\documentclass[reprint,aps]{revtex4-1}

\usepackage{amsmath, amsthm, amssymb, amsfonts}

\usepackage[utf8]{inputenc} 
\usepackage[T1]{fontenc}    
\usepackage{url}            
\usepackage{booktabs}       
\usepackage{nicefrac}       
\usepackage{microtype}      
\usepackage{graphicx}
\usepackage{graphics}
\usepackage{dcolumn}
\usepackage{bm}
\usepackage{upgreek}
\usepackage{latexsym}
\usepackage{xr}
\usepackage{mathptmx}
\usepackage[compact]{titlesec}
\usepackage{notoccite}

\begin{document}

\preprint{APS/123-QED}

\title{Influence of Surface Hydrophilicity and Hydration on the Rotational Relaxation of Supercooled Water on Graphene Oxide Surfaces}

\author{Rajasekaran M}
\affiliation{
Department of Chemical Engineering, Indian Institute of Science, Bangalore, India 560012
}
\author{ K. Ganapathy Ayappa}
\email{ayappa@iisc.ac.in}
\altaffiliation{
Centre for Biosystems Science and Engineering, Indian Institute of Science, Bangalore, India 560012
}
\affiliation{
Department of Chemical Engineering, Indian Institute of Science, Bangalore, India 560012
}
  	
\begin{abstract}
Hydration or interfacial water present in  biomolecules and inorganic solids have been shown to exhibit a dynamical transition upon supercooling. However, an understanding of the extent of the underlying surface hydrophilicity as well as the local distribution of hydrophilic/hydrophobic patches on the dynamical transition is unexplored. Here, we use molecular dynamics simulations with a TIP4P/2005 water model to study translational and rotational relaxation dynamics of interfacial water on graphene surfaces. The purpose of this study is to investigate the influence of both surface chemistry as well as the extent of hydration on the rotational transitions of interfacial water on  graphene oxide (GO) surfaces in the deeply supercooled region. We have considered three graphene-based surfaces; a GO surface with equal proportions of oxidized and pristine graphene regions in a striped topology, a fully oxidized  surface and a pristine graphene surface. The dipole relaxation time of interfacial water (high hydration) shows a strong-to-strong, strong, and a fragile-to-strong transition on these surfaces, respectively, in the temperature range of 210-298 K. In contrast, bulk water shows a fragile-to-strong rotational transition upon supercooling. In all these cases at high hydration, interfacial water co-exists with a thick water film with bulk-like properties. To investigate the influence of bulk water on dynamical transitions, we simulated a low hydration regime where only bound water (surface water) is present on the GO surfaces and found that the rotational relaxation of surface water on both the GO and fully oxidized surfaces show a single Arrhenius temperature dependence. Bulk water is found to have a greater influence on the rotational relaxation in the presence of a hydrophobic surface and the dipole angular distributions show distinct differences on the surfaces upon supercooling. 
Our results indicate that not only does the local extent of surface hydrophilicity play a role in determining the energy landscape explored by the water molecules upon supercooling, but the presence of bulk water also modulates the dynamic transition. 
\end{abstract}		

\maketitle	
						
\section{Introduction}
The presence of a surface is known to modulate the relaxation dynamics of interfacial  water upon supercooling.  Interfacial water in biomolecules, also known as hydration water, is essential for the activity and stability of biomolecules playing an important role in the conformational dynamics of proteins~\cite{levy2006water,ball2008water,zhong2011biological,bagchi2013water,bagchi2016untangling,laage2017water,ball2017water}.  Interfacial water has been extensively studied to probe the dynamics of supercooled water since bulk water cannot be experimentally probed below the homogeneous nucleation temperature, $T_{\text{H}}$ = 235 K and above 150 K where the amorphous, hyperquenched water  spontaneously crystallizes. In this so-called "no man's land" range of temperatures, interfacial water on surfaces and biomolecules have been shown to remain in the liquid phase~\cite{sartor1992calorimetric}, allowing one to study the dynamical transition of water upon supercooling.  It has been shown that the glass transition of proteins is strongly correlated with hydration water dynamics~\cite{Chen_2006,chen2006observation,kumar2006glass}. Although studying the dynamical transitions of confined or interfacial water to draw a connection with the thermodynamics of supercooled bulk water is an unsolved problem, the dynamics of interfacial water presents its own scientific challenges.  Understanding the dynamics of supercooled interfacial or hydration water will improve our understanding of technological relevant applications such as the elimination of hydrate formation in natural gas pipelines~\cite{debenedetti2003supercooled,ludwig2006puzzling} and cryopreservation of organs that are essential for life~\cite{debenedetti2003supercooled,franzese2010phase,ludwig2006puzzling}.

Several experimental and molecular dynamics simulations have been used to extensively study the nature of transitions in the hydration water of biomolecules such as DNA~\cite{chen2006observation, Chen_2006}, RNA \cite{chu2008observation,biswal2009dynamical} and  lysozyme~\cite{lagi2008low,kumar2006glass}. These studies reveal that the hydration water in biomolecules undergoes a dynamic slowdown upon supercooling with the relaxation times typically displaying a fragile-to-strong (FTS) transition upon supercooling. Although hydration water has been extensively studied in the context of biomolecules, water is ubiquitously present as hydration or interfacial water in a wide variety of solid surfaces, ranging from microporous materials such as zeolites to layered structures such as clays and graphene oxide membranes. Interfacial water on cerium oxide~\cite{Mamontov_2005} was found to display an FTS dynamic crossover at $T_{\text{c}}$ = 215 K. In contrast, the inner layer of surface water on zirconium oxide~\cite{mamontov2005high} showed a non-Arrhenius temperature dependence of the relaxation time in the temperature range of 240 - 300 K, and a crossover phenomenon was not observed. Cerveny et~al.~\cite{cerveny2010dynamics} carried out dielectric spectroscopy studies on confined water in graphite oxide and observed a non-Arrhenius to an Arrhenius crossover (weak transition) in dipole relaxation at 192 K. Quasielastic neutron scattering (QENS) studies~\cite{mamontov2006dynamics,chu2007observation} on water confined in the 14 \AA\, and 16 \AA\, carbon nanotubes (CNT) revealed a strong FTS transition at 218 K and 190 K, respectively. In experiments where both bulk and interfacial water are present, it is not always easy to separate the response of hydration water alone. 

The dynamics of interfacial water at hydrophobic and hydrophilic solid surfaces, such as magnesium oxide, alumina, silica, mica, and graphite, has been investigated using MD simulations~\cite{argyris2011molecular,argyris2011structure,skelton2011simulations,ho2011interfacial,deshmukh2012atomic,phan2012molecular,ho2013polarizability,ho2014molecular,huang2014alumina,monroe2018computational}. Furthermore, the translational and rotational dynamics of hydration water of lipid bilayers~\cite{senapati2003water,bhide2006behavior,debnath2010entropy} and proteins~\cite{zhang2009protein,mondal2017origin} have also been studied using MD simulations and simulations reveal that the dynamics of hydration water is slowed down when compared with that of bulk water. The interfacial water on reverse micelles exhibits a slow reorientation due to the strong influence of polar head groups in the reverse micelles~\cite{fenn2009water,fenn2011dynamics,biswas2012non}. In contrast to surfaces that are purely hydrophilic or hydrophobic with a uniform surface energy landscape, graphene oxide (GO) surfaces are particularly attractive, since the extent of hydrophobicity/hydrophilicity can be controlled and tuned for a given application during the synthesis of these unique atomically thin materials. With their nanoporous architecture and increased hydrophilicity when compared with graphite, GO membranes are widely investigated for their potential use in water desalination~\cite{joshi2014precise,raghav2015molecular,abraham2017tunable} and proton transport applications~\cite{karim2013graphene}. Although numerous molecular simulation studies have focused on the translational and rotational dynamics of water confined in GO membranes~\cite{wei2014understanding,raghav2015molecular,devanathan2016molecular,willcox2017molecular2,dai2016water,raja2019enhanced}, water dynamics in the supercooled regime has not been investigated. 

In this manuscript, we use molecular dynamics simulations to study the relaxation dynamics of interfacial water present in graphene oxide (GO) nanosheets. Graphene oxide nanosheets are widely studied due to their atomic thickness, ability to functionalize biomolecules, and unique properties due to the presence of hydrophilic and hydrophobic interfaces that can be tuned for a given application. Thus, the graphene oxide nanosheet is an ideal candidate to study the influence of hydrophilicity on supercooled interfacial water dynamics. We probe the dynamics of water present in the first hydration layer both in the presence of a bulk water film adjacent to the GO nanosheet as well as a low hydration situation where only bound or surface water is present. In what follows we classify and refer to these as {\it interfacial and surface water} respectively. We study the dynamics of supercooled interfacial water on graphene oxide nanosheets  at temperatures ranging from 298 - 210 K using molecular dynamics simulations with the TIP4P/2005 water model~\cite{abascal2005general}, which is known to accurately capture several key properties of bulk water.  Two types of surfaces are contrasted. In one case, the surface is striped, containing both hydrophobic and hydrophilic parts, which we will refer to as the GO surface. In the other cases, the surface is either fully oxidized (the O surface) giving rise to a hydrophilic surface, or pristine carbon giving rise to a graphene (G) surface.  The GO surfaces used in this study represent a typical interface that a water molecule encounters in GO membranes~\cite{willcox2017molecular2,raja2019enhanced}. We aim to determine the influence of surface oxidation as well as the level of hydration on the dynamic crossover of supercooled interfacial water by evaluating the water structure, mean squared displacements, the  dipole-dipole correlation function and the rotational relaxation times. The temperature dependence of the rotational relaxation times is used to classify the interfacial water as either fragile or strong and illustrate the influence of surface chemistry on the nature of the dynamical transition upon supercooling.

\section{Molecular dynamics simulation details}

\begin{figure*}[!htpb]
	\centering
	\includegraphics[scale=0.45]{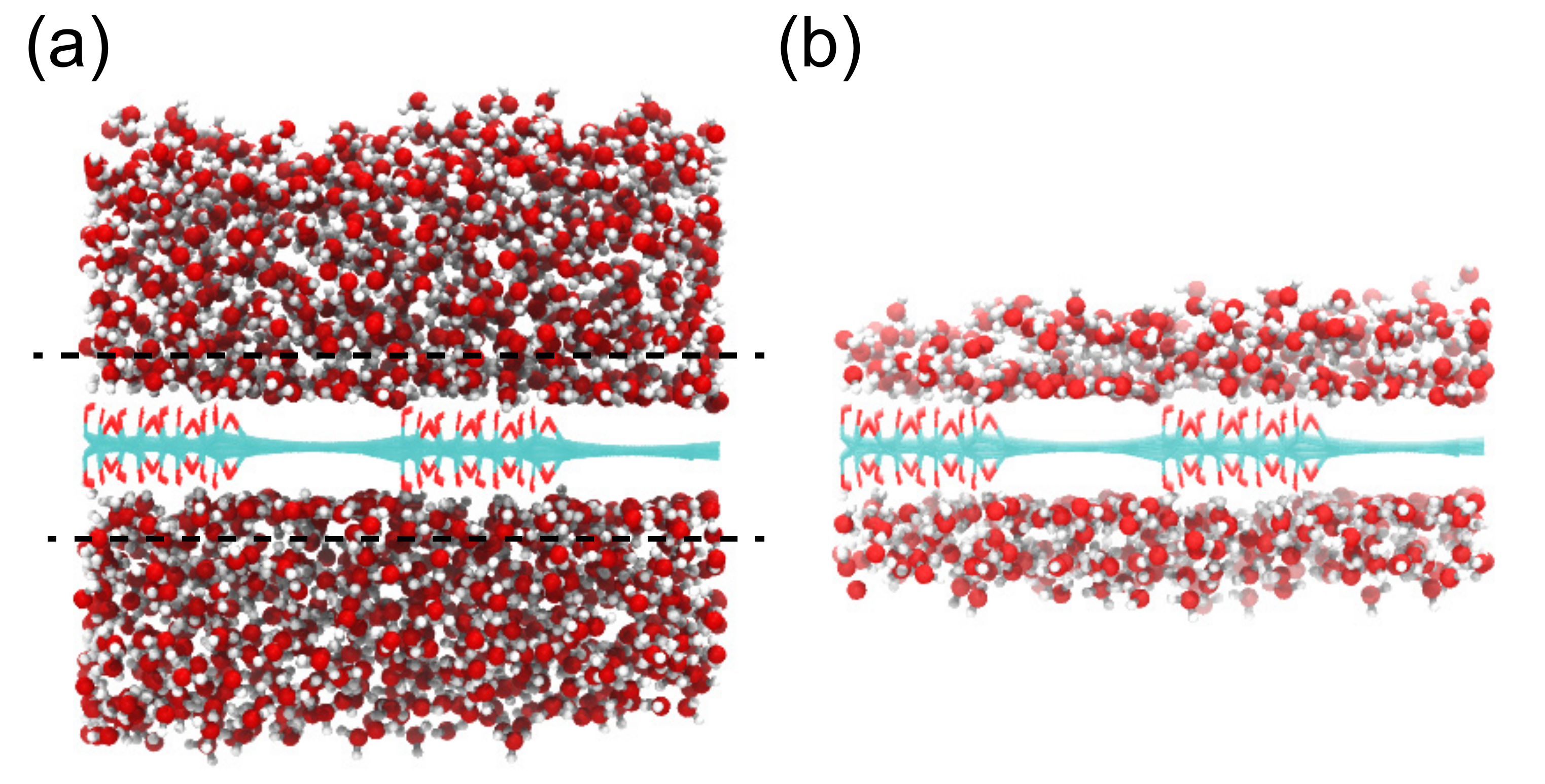}
	\caption[Representative molecular snapshots of supercooled interfacial water]{Representative molecular snapshots of (a) interfacial water molecules on the GO surface and (b) surface water on the GO surface. The dashed line indicates the distance from the surface within which the dynamics of water at high hydration levels is analyzed. This represents water contained in the first contact layer (Fig.~\ref{density_GO_LL}). The size of the simulation box is L$_{x}$ $\approx$ 102 \AA\,, L$_{y}$ $\approx$ 89 \AA\, and L$_{z}$ = 200 \AA\,. The number of water molecules in case of high hydration and low hydration is 8000 and 2270, respectively.}
	\label{snap_IW}
\end{figure*}

Molecular dynamics (MD) simulations were performed to investigate the dynamics of supercooled interfacial water on three graphene-based surfaces in the canonical ($NVT$) ensemble using LAMMPS~\cite{plimpton1995fast}. The simulation box consists of a surface of L$_{x}$ $\approx$ 102 \AA\,, L$_{y}$ $\approx$ 89 \AA\,placed at the center of the box with $\sim$ 8,000 water molecules for high hydration (Fig.~\ref{snap_IW}a) and 2,270 water molecules in the case of low hydration (Fig.~\ref{snap_IW}b). Periodic boundary conditions were imposed in the $x$ and $y$ directions with a vacuum above and below the surface in the $z$ - direction. The periodic box dimensions including the vacuum region are L$_{x}$ $\approx$ 102 \AA\,, L$_{y}$ $\approx$ 89 \AA\,, and L$_{z}$ = 200 \AA\,. The TIP4P/2005 water model~\cite{abascal2005general}, which is well-known for reproducing dynamical properties, and the phase diagram of water, was used for the water potential with the SHAKE algorithm for restraining the bond angle and bond lengths of a water molecule. Three different graphene-based surfaces (graphene surface with strips of oxidized regions (GO), completely oxidized graphene (O) surface, and graphene (G) surface) were considered to study the influence of surface chemistry on the dynamics of water. The GO and O surfaces were optimized by performing periodic density functional theory calculations using Quantum ESPRESSO~\cite{QE}. The partial atomic charges of the GO and O surfaces were computed with the CHELPG scheme at the level of the Hartree-Fock/6-31G* basis set using the Gaussian software~\cite{g09}. The optimized coordinates and partial atomic charges of atoms on the GO surfaces were similar to those used in an earlier study~\cite{raja2019enhanced}. All surfaces were kept rigid by fixing the positions of all the atoms of the surface. All-atom optimized potentials for liquid simulation (OPLS-AA) parameters were employed for the graphene oxide surfaces along with computed charges~\cite{shih2011understanding,willcox2017molecular2,wei2014breakdown,wei2014understanding,willcox2017molecular}. 

The Lennard-Jones (LJ) 12-6 potential was used to model the interactions between GO surfaces and water, with the Lorentz-Berthelot mixing rules. The cutoff for the LJ interactions was set to 12 \AA\,. The particle-particle particle-mesh (PPPM) algorithm~\cite{hockney1980computer} was used to compute the long-range electrostatic interactions. The velocity Verlet scheme was employed to integrate the equations of motion with a time step of 2 fs. The Nos$\acute{e}$-Hoover thermostat with a time constant of 0.1 ps was employed to maintain the system temperature. Simulations were carried out by sequentially cooling the system with equilibration for 10 - 40 ns depending on the temperature, followed by a 5 ns production run, during which the properties are computed. Correlation functions and mean squared displacements were computed for water molecules that reside continuously in the region of interest. Water residence times were found to decay rapidly within a 5 ns time duration which was sufficient to analyze dynamical properties (Table~\ref{dipole_SE_IW}). The trajectories of water molecules were stored at every 4 - 20 fs to analyze the structural and dynamical properties.
 
\section{Results}

\subsection{High Hydration: Interfacial Water}
\subsubsection{Density distributions}
\begin{figure*}[!htpb]
	\centering
	\includegraphics[scale=0.75]{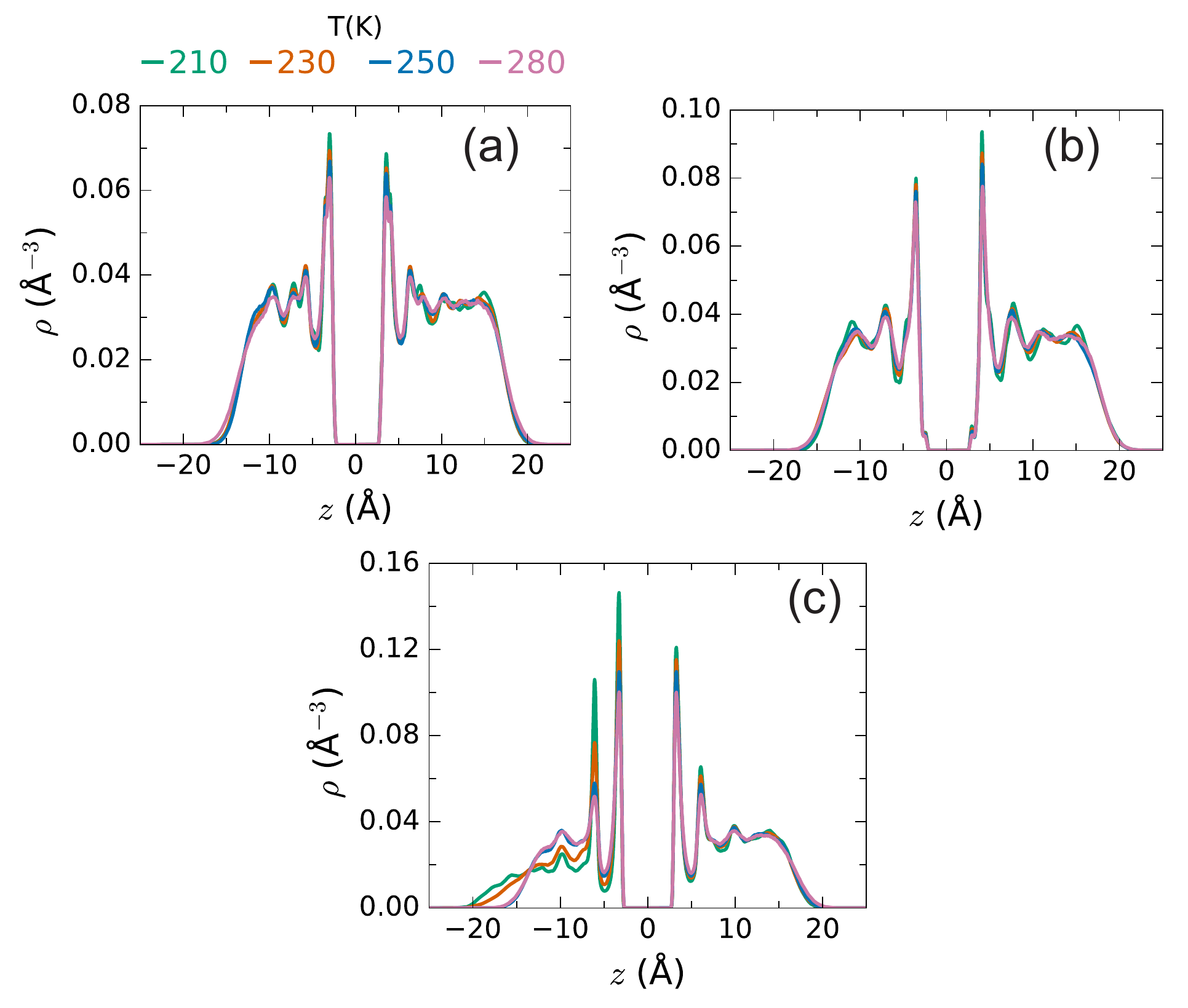}
	\caption[Density distributions of supercooled interfacial water]{Density distributions of water molecules normal to (a) GO surface, (b) O surface, and (c) G surface at different temperatures}
	\label{density_GO_LL}
\end{figure*}

We study the layering of water on the surface by computing the layer-averaged density along the surface normal using,

\begin{equation} \label{density}\nonumber
\rho(z) = \frac{\bigg\langle N(z-\frac{\Delta z}{2},z+\frac{\Delta z}{2})\bigg\rangle}{A\Delta z}   ,
\end{equation}

where $N(z-\frac{\Delta z}{2},z+\frac{\Delta z}{2})$ is the number of water molecules in a bin of thickness $\Delta z$ in the $z$ direction, $A = L_{x}L_{y}$ , and $\langle .. \rangle $ denotes a time average. In Fig.~\ref{density_GO_LL}a, we show the density distributions of water molecules in the $z$-direction (surface normal) for the GO, O, and G surfaces at different temperatures. The density profile on the GO surface as illustrated in Fig.~\ref{density_GO_LL}a reveals a $\sim$ 3 \AA\, thick distinct contact water layer on both the sides ($z$ > 0 and $z$ < 0), followed by a second layer and oscillations extending up to 15 \AA\, from the GO surface. The density peak of the first layer is significantly higher than that of bulk water (0.033 \AA$^{-3}$), and the extended density oscillations are due to the influence of the GO surface. We observe a 3.5 \AA\, thick distinct water layer on both sides of the O surface, as shown in Fig.~\ref{density_GO_LL}b. The peak intensity of the water layer on the O surface is higher than that on the GO surface. 

We observe an asymmetry in the density distributions on either side of the GO sheet and attribute this to the inherent structural differences present on the two surfaces~\cite{raja2019enhanced}. The density peak on both surfaces increases on decreasing temperature. However, the width of the first water layer on both the surfaces remains unchanged with a decrease in temperature. The graphene (G) surface exhibits a 2.75 \AA\, thick distinct first water layer with the peak intensity higher than that of GO and O surfaces.  As shown in Fig.~\ref{density_GO_LL},  water molecules present in the first contact layer on the GO, O, and G surfaces are considered as interfacial water and used for further analysis.

It is useful to compare the layering of water observed with other hydrophilic surfaces such as silica and mica. In the case of mica, the first contact layer has a lower density when compared with the second adsorbed layer, since water molecules adsorb onto the vacant sites unoccupied by potassium on the mica surface in the first contact layer~\cite{malani2009adsorption}. The density distributions on silica is a strong function of the extent of hydroxylation~\cite{argyris2009dynamic}, and the density peak in partially hydroxylated silica surface is significantly higher than the GO surface.


\subsubsection{In-plane pair correlation function}

\begin{figure*}[!htpb]
	\centering
	\includegraphics[scale=0.75]{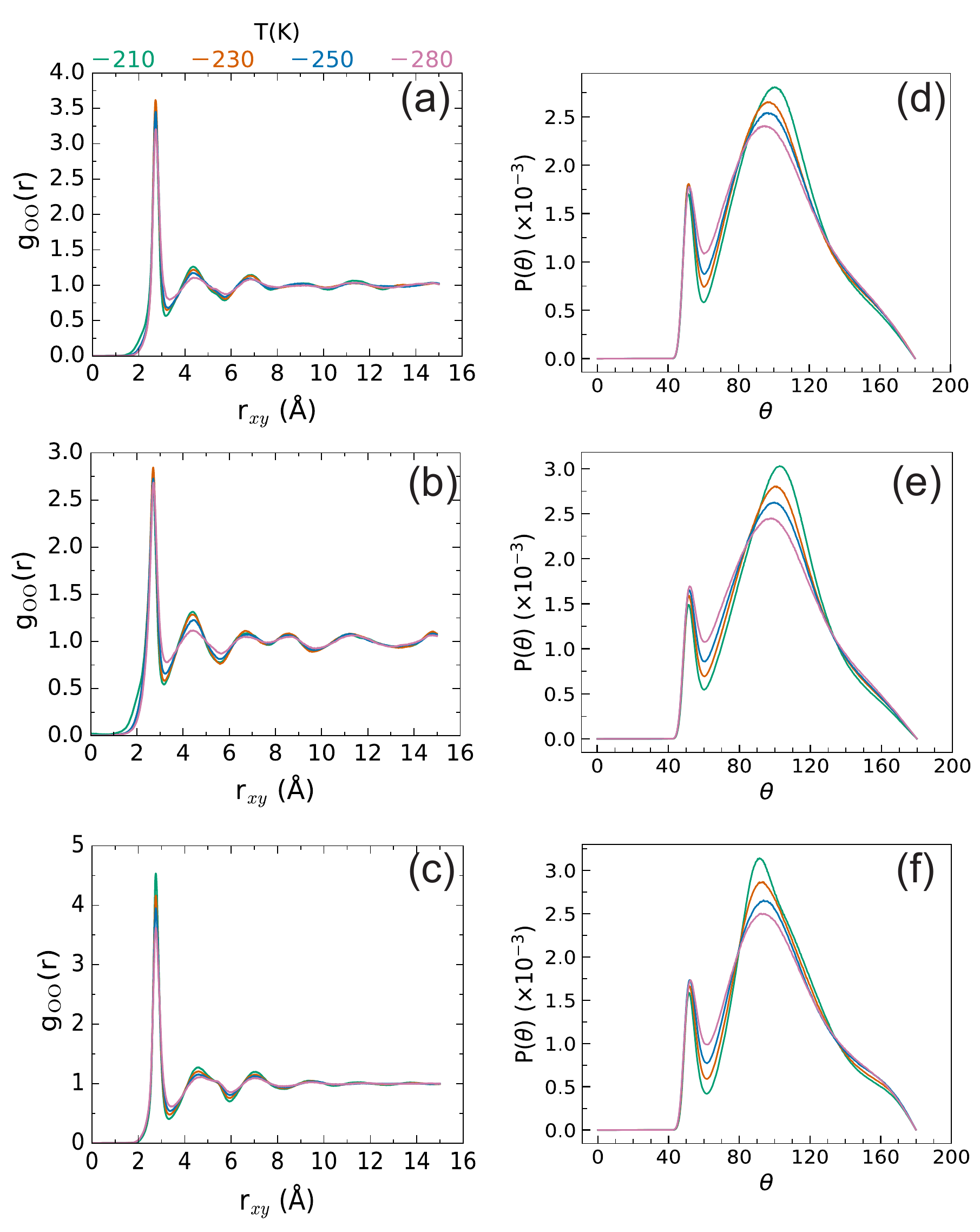}
	\caption[Structure of supercooled interfacial water]{In-plane oxygen-oxygen pair correlation function of interfacial water on (a) GO surface, (b) O surface, and (c) G surface, Oxygen-oxygen-oxygen angle distribution of the interfacial water molecules in the first contact layer on (d) GO surface (e) O surface, and (f) G surface at various temperatures }
	\label{rdf_IW}
\end{figure*}

In order to study the local structure of the interfacial water molecules on the surfaces, we compute the in-plane oxygen-oxygen pair correlation function using,
\begin{equation} \label{2drdf}\nonumber
g_{ij}(r) = \Bigg \langle \sum\limits_{i=1}^{N_L} \sum\limits_{j=1}^{N_L} \frac{N_{ij}(r+\frac{\Delta r}{2},r - \frac{\Delta r}{2}) A}{2 \pi r \Delta r N_{i} N_{j}} \Bigg \rangle ,
\end{equation}
where $ N_{ij}(r+\frac{\Delta r}{2},r - \frac{\Delta r}{2})$ is the number of atoms $j$  around a atom $i$ at a distance $r$ in a cylindrical shell of thickness $\Delta r$. $N_L$ is the number of particles in the first contact layer adjacent to the surface, and the boundaries of the first layer are obtained from the density distributions (Fig.~\ref{density_GO_LL}). Figures~\ref{rdf_IW}a-c show the in-plane oxygen-oxygen pair correlation of the interfacial water on the GO, O and G surfaces at various temperatures. The interfacial water on the O surface, as shown in Fig.~\ref{rdf_IW}b, shows largest oscillations, indicating the presence of increased in-plane ordering when compared with either the GO or G surface. These oscillations are suppressed due to the pristine graphene in the GO surface, which disrupts the hydrogen bonding for the interfacial water. The pair correlation function of interfacial water on the graphene (G) surface also exhibits increased oscillations at lower temperatures (Fig.~\ref{rdf_IW}c) although to a lesser extent when compared with the other surfaces. 

\subsubsection{Angle distribution of oxygen-oxygen-oxygen triplet}

The oxygen-oxygen-oxygen triplet angle has been used as a measure of the tetrahedral arrangement of interfacial water ~\cite{malaspina2010structural,towey2016low,biswal2009dynamical}. The triplet angle is computed only when the O-O distances are less than 3.55 \AA\,.  For a perfectly tetrahedral arrangement, a peak should occur at an angle of 109.5$^\circ$. In Figs.~\ref{rdf_IW}d-f,  the angular distribution of oxygen-oxygen-oxygen triplets of interfacial water show the presence of an additional smaller peak at 50$^{\circ}$. This has been attributed to water with reduced hydrogen bonding~\cite{biswal2009dynamical}, present in the first hydration shell of an oxygen atom and a weak decrease in the peak intensity is observed with decreasing temperature. The broad second peak at around 100$^{\circ}$ indicates the distorted local tetrahedral arrangement of interfacial water present on the GO surface, and its intensity increases with decreasing temperature due to increased hydrogen bonding. Interestingly, the greatest increase in the peak around 100$^{\circ}$ was observed for the G surface (Fig.~\ref{rdf_IW}f) with the maxima occurring at $T$ = 210 K. In contrast the peaks for the O and G surfaces was found to shift toward higher angles at reduced temperature. This suggests that water on the G surface is able to reorient  and maximize hydrogen bonding to a greater extent with surrounding water molecules when compared with the other surfaces.


\subsubsection{In-plane diffusion}

\begin{figure*}[!htpb]
	\centering
	\includegraphics[scale=0.75]{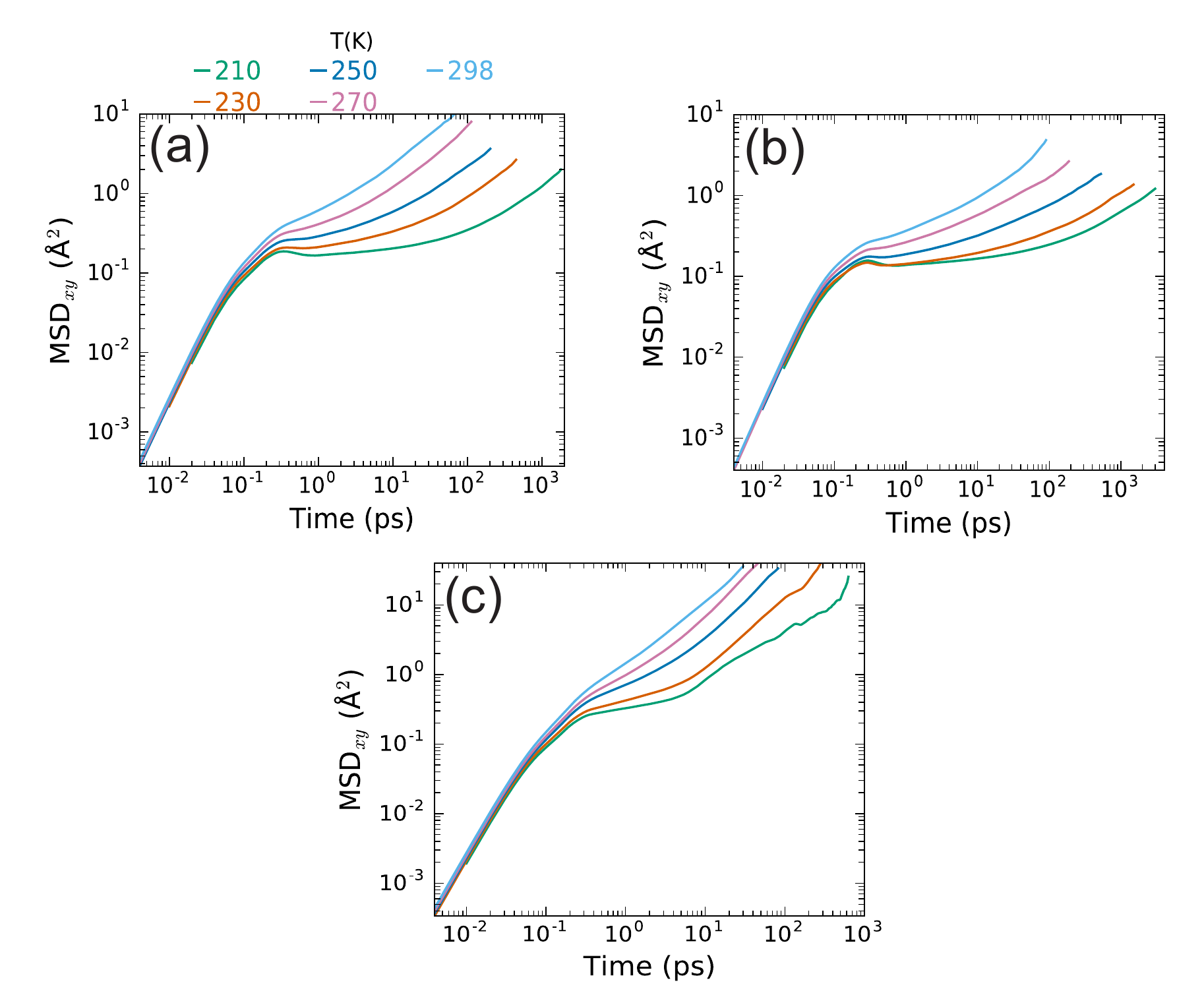}
	\caption[Translational dynamics of supercooled interfacial water]{MSD$_{xy}$ of interfacial water molecules on the (a) GO surface, (b) O surface, and (c) G surface at various temperatures }
	\label{msd_IW}
\end{figure*}

The translational dynamics of interfacial water is studied by analyzing in-plane mean-squared displacement (MSD$_{xy}$) using,

\begin{equation}\label{msd_2d}\nonumber
{\mbox {MSD}}_{xy}(t)  = \Bigg\langle \frac{1}{N}\sum\limits_{i=1}^{N} \vert {x}_{i}(t)-{x}_{i}(0)\vert ^{2} + \vert{y}_{i}(t)-{y}_{i}(0)\vert^{2} \Bigg\rangle_{\tau} ,
\end{equation}

where ${x}_{i}(t) $ and ${y}_{i}(t)$ are the $x$ and $y$ coordinates of the center of mass of molecule $i$, $N$ is the total number of molecules, and $\langle ... \rangle _{\tau}$ is the ensemble average over $\tau$ shifted time origins. Only interfacial water molecules continuously surviving in the first contact layer are used for the computation of MSD$_{xy}$. In Fig.~\ref{msd_IW}a-c, we illustrate the time dependence of MSD$_{xy}$ of interfacial water on the GO, O, and G surfaces at different temperatures. The ballistic regime, which is the collision-free region where the MSD$_{xy}$ scales as $t^{2}$ at short times, is observed for all surfaces at all temperatures. The weak subdiffusive regime where the MSD$_{xy}$ scales as $t^{\alpha}$ with $\alpha$ < 1.0 at intermediate times for GO and O surfaces at all temperatures. At lower temperatures, the weak, short-lived flattening in the MSD$_{xy}$,  which is indicative of caging, is observed for the interfacial water molecules on the GO and O surfaces. Since interfacial water molecules did not exhibit diffusive behavior within the residence times in the
the layer, in-plane self-diffusion coefficients were not evaluated. 
%

\subsubsection{Rotational dynamics}

\begin{figure*}[!htpb]
	\centering
	\includegraphics[scale=0.75]{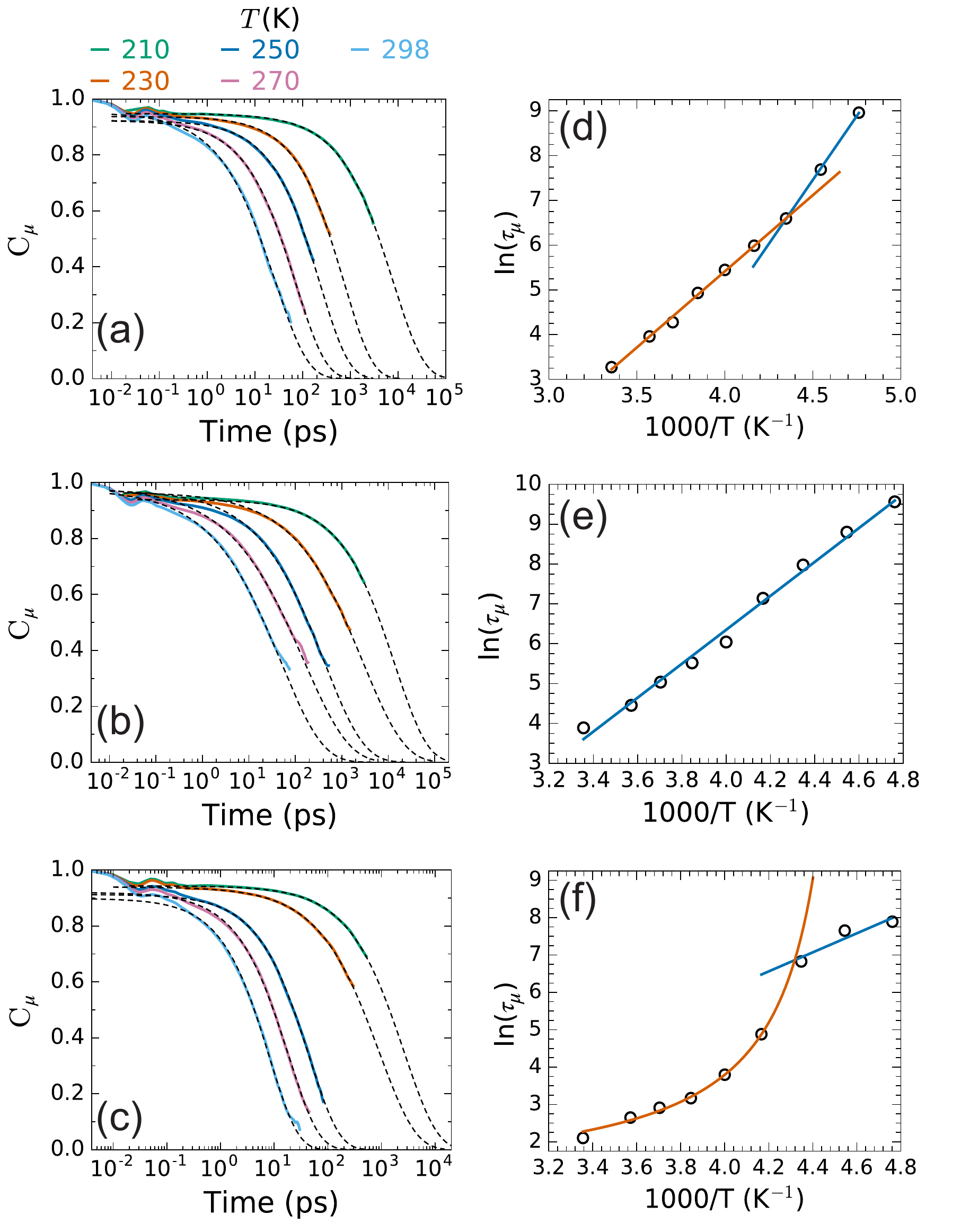}
	\caption[Dynamic crossover in dipole relaxation time of supercooled interfacial water]{Dipole-dipole correlation function of interfacial water molecules on (a) GO surface (b) O surface, and (c) G surface at various temperatures. The continuous lines are the MD simulation results, whereas dashed lines are fits obtained from Eq.~\ref{fit_ocf_IW}. The model parameter obtained from the fits of $C_{\mu}$ are given in Table~\ref{dipole_SE_IW}. Temperature dependence of the dipole relaxation time of interfacial water molecules on the (d) GO surface, (e) O surface, and (f) G surface. The open circles are simulation data and the solid lines are fits.}
	\label{DCF_IW}
\end{figure*}

In order to analyze the rotational dynamics of interfacial water molecules, we compute the dipole-dipole time correlation function using,

\begin{equation} \label{ocf} \nonumber
C_{\mu}(t) = \Bigg \langle \frac{1}{N} \sum\limits_{i=1}^{N} (\textbf{e}^{\mu}_{i}(t) \cdot \textbf{e}^{\mu}_{i}(0)) \Bigg \rangle_{\tau},
\end{equation}

where $\textbf{e}^{\mu}_{i}$ is the dipole moment unit vector of water molecule $i$ in the molecular frame. Figure~\ref{DCF_IW}a-c shows the time dependence of the dipole-dipole correlation function of interfacial water on the GO, O, and G surfaces at different temperatures. The dipole-dipole correlation function, $C_{\mu}(t)$, is computed for only molecules that continuously reside in the layer up to time $t$. We use a stretched exponential function~\cite{kumar2006molecular,biswal2009dynamical} to model the $C_{\mu}(t)$.

\begin{equation}\label{fit_ocf_IW}
C_{\mu}(t) =  a\exp \left[-(t/\uptau_{\mu})^\beta \right], 
\end{equation} 
   
where $a$ is a constant, $\beta$ is the stretched exponent ranging from 0 to 1, and $\uptau_{\mu}$ is the dipole relaxation time. A simple exponential function is recovered when the stretched exponent is 1. The model parameters for the $C_{\mu}$ data are provided in Table~\ref{dipole_SE_IW}. The dipole-dipole correlation function, $C_{\mu}$ displays a fast relaxation at short times, followed by a slow relaxation at long times. This dipole correlation function does not decay to zero within the time window considered since molecules do not continuously reside within the layer to contribute to the long-time data. The extrapolated fits from Eq.~\ref{fit_ocf_IW} illustrate the time evolution at longer times. As the temperature is reduced, the long-time relaxation becomes increasingly slow. The relaxation time is greatest for the O surface and least for the G surface.  The stretching exponent $\beta$ is the least for the O surface indicating that water experiences the greatest degree of rotational heterogeneity when adjacent to a fully oxidized surface which increases the hydrogen bonding with the surface. Values of $\beta$ are greatest for the G surface, which is purely hydrophobic, followed by the values for the GO surface.

\begin{table} [!htpb]
	\caption[Model parameters for the dipole correlation function of supercooled interfacial water]{Model parameters obtained from the fitting of $C_{\mu}$ of interfacial water molecules on the GO, O, and G surfaces at various temperatures. The $C_{\mu}$ is fitted to Eq.~\ref{fit_ocf_IW}.} 
		\centering
	\begin{ruledtabular}
		\begin{tabular} { c c c c c c c }
		Surface & T(K) & t$_{w}$ (ps) & N$_{w}$ & a  & $\beta$ & $\uptau_{\mu}$ (ps)  \\	 
		\hline		
		GO  & 210 & 4000 & 263& 0.9478 & 0.6706 & 7804 \\
		& 220 &  1000 & 217 & 0.9372 & 0.7618 & 2184 \\
		& 230 &  500 & 216 & 0.9373 & 0.7299 & 732.9 \\
		& 240 &  400 & 186 & 0.925  & 0.7358 & 398.4  \\
		& 250 &  225 & 232 & 0.923  & 0.7178 & 232 \\
		& 260 &  180 & 194 & 0.9316 & 0.6686 & 138.7  \\
		& 270 &  150 & 172 & 0.9232 & 0.6935 & 72.07 \\
		& 280 &  80 & 213 & 0.9587 & 0.6047 & 52.46  \\
		& 298 &  75 & 131 & 0.952  & 0.6291 & 26.38   \\		
		\\
		O  & 210 & 4000 & 822 & 0.9392  & 0.6271 & 14200 \\
		& 220 &  2000 & 473 & 0.9495 & 0.551  & 6628  \\
		& 230 &  2000 & 222 & 0.9663 & 0.4886 & 2907 \\
		& 240 &  1000 & 230 & 0.994  & 0.4269 & 1261 \\
		& 250 &  700 & 187 & 0.9741 & 0.5068 & 420.8 \\
		& 260 &  400 & 200 & 0.9464 & 0.5583 & 249.5  \\
		& 270 &  250 & 198 & 0.9846 & 0.4509 & 153.9 \\
		& 280 &  120 & 316 & 0.9827 & 0.4941 & 85.94  \\
		& 298 &  100 & 186 & 0.9768 & 0.4834 & 48.86   \\            
		\\
		G  & 210 &  700 & 646 & 0.9433  & 0.7105  & 2675 \\
		& 220 &  600 & 264 & 0.9412  & 0.6314  & 2108  \\
		& 230 &  400 & 205 & 0.9394  & 0.6474  & 923.2 \\
		& 240 &  175 & 206 & 0.9315  & 0.6959  & 131.7  \\
		& 250 &  110 & 215 & 0.9125  & 0.7776  & 44.55 \\
		& 260 &  75 & 197 & 0.894   & 0.8976  & 23.78  \\
		& 270 &  60 & 168 & 0.9199  & 0.7731  & 18.4 \\
		& 280 &  50 & 162 & 0.8896  & 0.8622  & 14.14  \\
		& 298 &  40 & 108 & 0.8982  & 0.8208  & 8.193   \\
	\end{tabular}
	\end{ruledtabular}
		\footnote{t$_{w}$ denotes the time window for analysis and N$_{w}$ represents the number of water molecules continuously reside in the first contact layer during the time window.}
	\label{dipole_SE_IW}
\end{table}

\begin{figure*}[!htpb]
	\centering
	\includegraphics[scale=0.75]{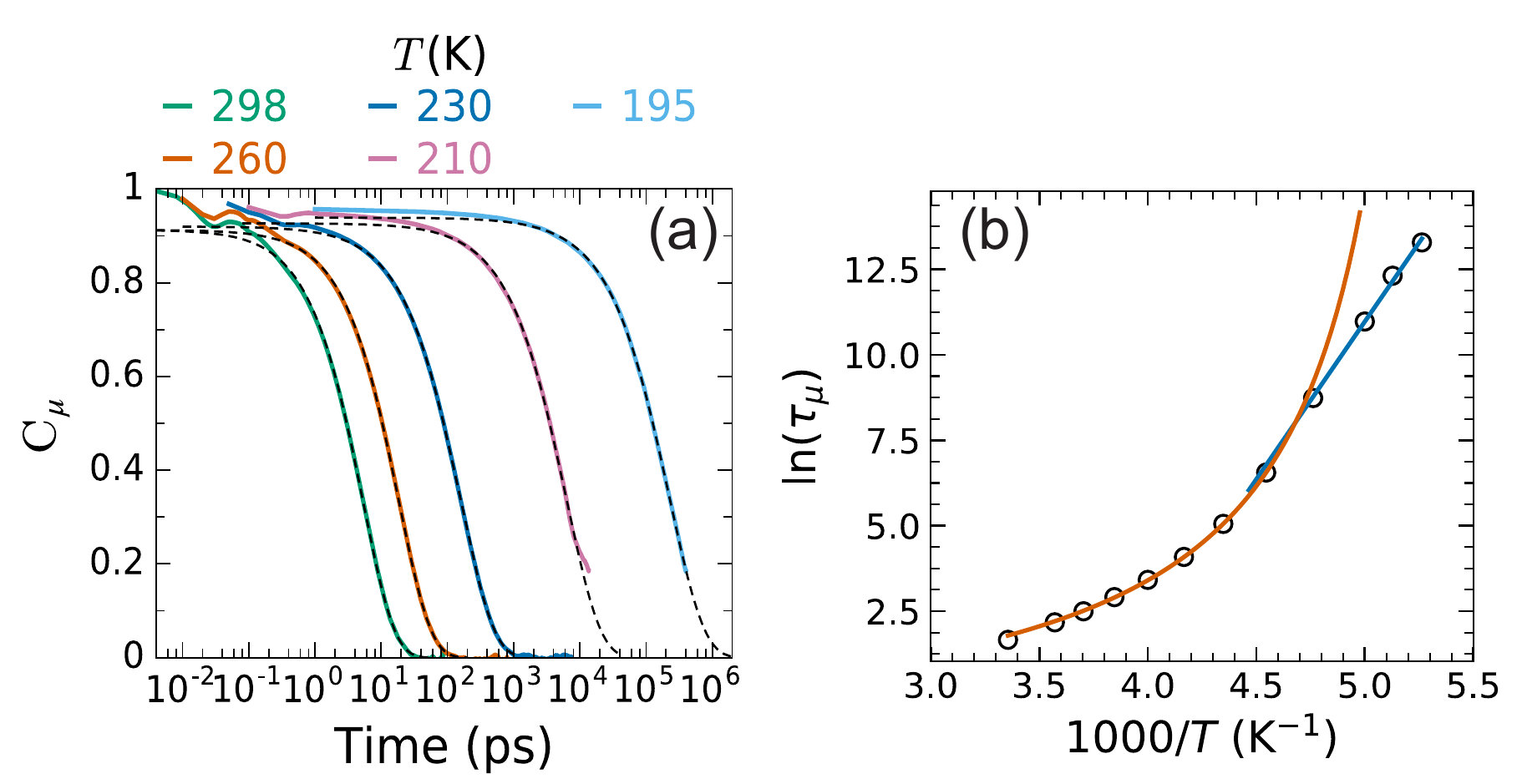}
	\caption[Dynamic crossover in dipole relaxation time of supercooled bulk TIP4P/2005 water]{(a) Dipole-dipole correlation function of bulk water at various temperatures. The continuous lines are the MD simulation results, whereas dashed lines are fits obtained from Eq.~\ref{fit_ocf_IW}. The model parameter obtained from the fits of $C_{\mu}$ are given in Table~\ref{dipole_SE_BW} (b) Temperature dependence of the dipole relaxation time of bulk water. The open circles are simulation data and the solid lines are fits.}
	\label{DCF_BW}
\end{figure*}

We fit the temperature dependence of the relaxation time to either an Arrhenius equation,
\begin{equation}\label{arrhenius}
\uptau_{\mu} = \uptau_{0}\exp(E_{\text{A}}/RT),
\end{equation}
where $E_{\text{A}}$ is the activation energy,  or the Vogel-Fulcher-Tammann (VFT) form,
\begin{equation}\label{vft}
\uptau_{\mu} = \uptau_{0}^{\text{VFT}}\exp(BT_{0}/(T-T_{0})), 
\end{equation}
where $B$ is the fragility parameter and $T_{0}$ is the ideal glass-transition temperature. 
Since the rotational relaxation dynamics for bulk TIP4P/2005 has not been reported, we also performed MD simulations of bulk water in a cubic box of 1000 water molecules described by the TIP4P/2005 potential for the temperature range of 298 - 190 K. NPT MD simulations were carried out at 1 atm pressure and different temperatures, followed by NVT simulations with corresponding densities obtained from previous NPT simulations. Shown in Fig.~\ref{DCF_BW}a is the bulk water dipole-dipole correlation function for selected temperatures, and these correlation functions are fitted to the stretched exponential function to obtain the dipole relaxation time. Table~\ref{dipole_SE_BW} summarizes the model parameters obtained from the fits of $C_{\mu}$. The dipole relaxation time data in the temperature range of 298 -220 K were fit to the VFT equation with $T_{0}$ = 183 K and $B$ = 1.4, and the low-temperature relaxation time data (210 - 190 K) fit well with the Arrhenius equation with $E_{A}$ = 76.68 kJ/mol. The fragile-to-strong dynamic transition in rotational relaxation is observed at $T_{C}$ = 214 K for the TIP4P/2005 water model (Fig.~\ref{DCF_BW}b). This dynamic transition temperature is lower than the freezing point of the TIP4P/2005 water model ($T_{m}$ = 252 K~\cite{abascal2005general}). The ideal glass transition temperature and fragility parameter are consistent with the results of Saito et~al.~\cite{saito2018crucial} where the translational relaxation dynamics of TIP4P/2005 water was studied. In contrast to the distinct FTS transition observed for the rotational relaxation, Saito et~al.~\cite{saito2018crucial} observed a transition from a fragile to a weak non-Arrhenius behavior above and below 220 K, respectively.

\begin{table}[!htpb]	
	\caption[Model parameters for the dipole correlation function of supercooled bulk water]{Model parameters obtained from the fitting (Eq.~\ref{fit_ocf_IW}) of $C_{\mu}$ of bulk water at various temperatures.} 
	\centering
	\begin{ruledtabular}
	\begin{tabular} { c c c c }
		T(K) &  a  & $\beta$ & $\uptau_{\mu}$ (ps)  \\
				\hline	 		
		190 &  0.9816 & 0.5932 & 588700 \\
		195 &  0.9389 & 0.8195 & 222300 \\
		200 &  0.9199 & 0.81 & 58260 \\
		210 &  0.9268 & 0.8436 & 6212 \\
		220 &  0.9308 & 0.813 & 703.2 \\
		230 &  0.9199 & 0.8517 & 156.6 \\
		240 &  0.9074 & 0.9 & 59.55  \\
		250 &  0.9167  & 0.8834 & 30.45 \\
		260 &  0.9126 & 0.896 & 18.38  \\
		270 &  0.9071 & 0.9221 & 12.15 \\
		280 &  0.9103 & 0.907 & 8.795  \\
		298 &  0.9129  & 0.9102 & 5.271   \\								
	\end{tabular}
	\end{ruledtabular}
	\label{dipole_SE_BW}
\end{table}

Figure~\ref{DCF_IW}d-f illustrates the temperature dependence of the dipole relaxation time of interfacial water molecules on different surfaces. For the interfacial water on the GO surface (Fig.~\ref{DCF_IW}d), the dipole relaxation time fits well to an Arrhenius form (Eq.~\ref{arrhenius}) in both the high and low-temperature regions, indicative of the presence of a strong-to-strong (STS) transition. Two distinct regimes are present, giving rise to an activation energy, $E_{\text{A}}$ = 28.28 kJ/mol in the high temperature region (298 - 240 K) and $E_{\text{A}}$ = 47.51 kJ/mol in the low temperature region (230 - 210 K). The dynamic crossover of this strong-to-strong transition of the rotational relaxation occurs at $T_{\text{c}}$ = 230 K for the interfacial water on the GO surface (Fig.~\ref{DCF_IW}d). Biswal et~al. ~\cite{biswal2009dynamical} also reported a similar strong-to-strong dynamic crossover for the supercooled hydration water (TIP5P water model) in the major and minor grooves of the Dickerson dodecamer DNA duplex. In the same study, they reported that the dynamic crossover of rotational motion takes place at temperature $T_{\text{c}}$ = 255 K for both major and minor grooves. It is noteworthy that this crossover temperature is $\sim$ 20 K below the freezing temperature of TIP5P water used in their study. Our study also shows the dynamic crossover at 22 K below the freezing temperature of the TIP4P/2005 water model. 
Hydration water of DNA is exposed to a certain degree of confinement inside the biomolecule, whereas  the interfacial water on the GO surface is purely surface-induced. 

In contrast, the dipole relaxation time of interfacial water on the O surface possesses only a single activation energy over the entire temperature range (298 - 210 K) with $E_{\text{A}}$ = 35.39 kJ/mol and a dynamic crossover for the interfacial water on the O surface is not observed within this temperature range (Fig.~\ref{DCF_IW}e). The activation energy for  interfacial water on the O surface is found to lie between the two activation energies obtained for the GO surface. The relaxation behavior for the G surface is quite different (Fig.~\ref{DCF_IW}f), showing a distinct FTS transition. The high-temperature data (298 - 240 K) are fitted to the VFT equation (Eq.~\ref{vft}) with $B$ = 0.41 and $T_{0}$ = 216 K. This behavior is also apparent in the relaxation time data (Table~\ref{dipole_SE_IW}) where a rapid increase in the values of the relaxation time, $\uptau_{\mu}$ is apparent for the G surface when compared with values in the same temperature range (298 - 240 K) for the GO surface. In the low-temperature range (230 -210 K) for the G surface, the data are fitted to an Arrhenius form with $E_{\text{A}}$ = 21.17 kJ/mol  This activation energy is significantly lower than that of either the  GO or O surfaces. The FTS dynamic crossover occurs at  $T_{\text{c}}$ = 232 K for the interfacial water on the graphene surface. This crossover temperature is lower than the freezing point of 252 K for the TIP4P/2005 water model. Interestingly, the dynamic crossover takes place at the same temperature for interfacial water on both the GO and G surfaces.

\subsection{Low Hydration: Surface Water}

\begin{figure*}[!htpb]
	\centering
	\includegraphics[scale=0.75]{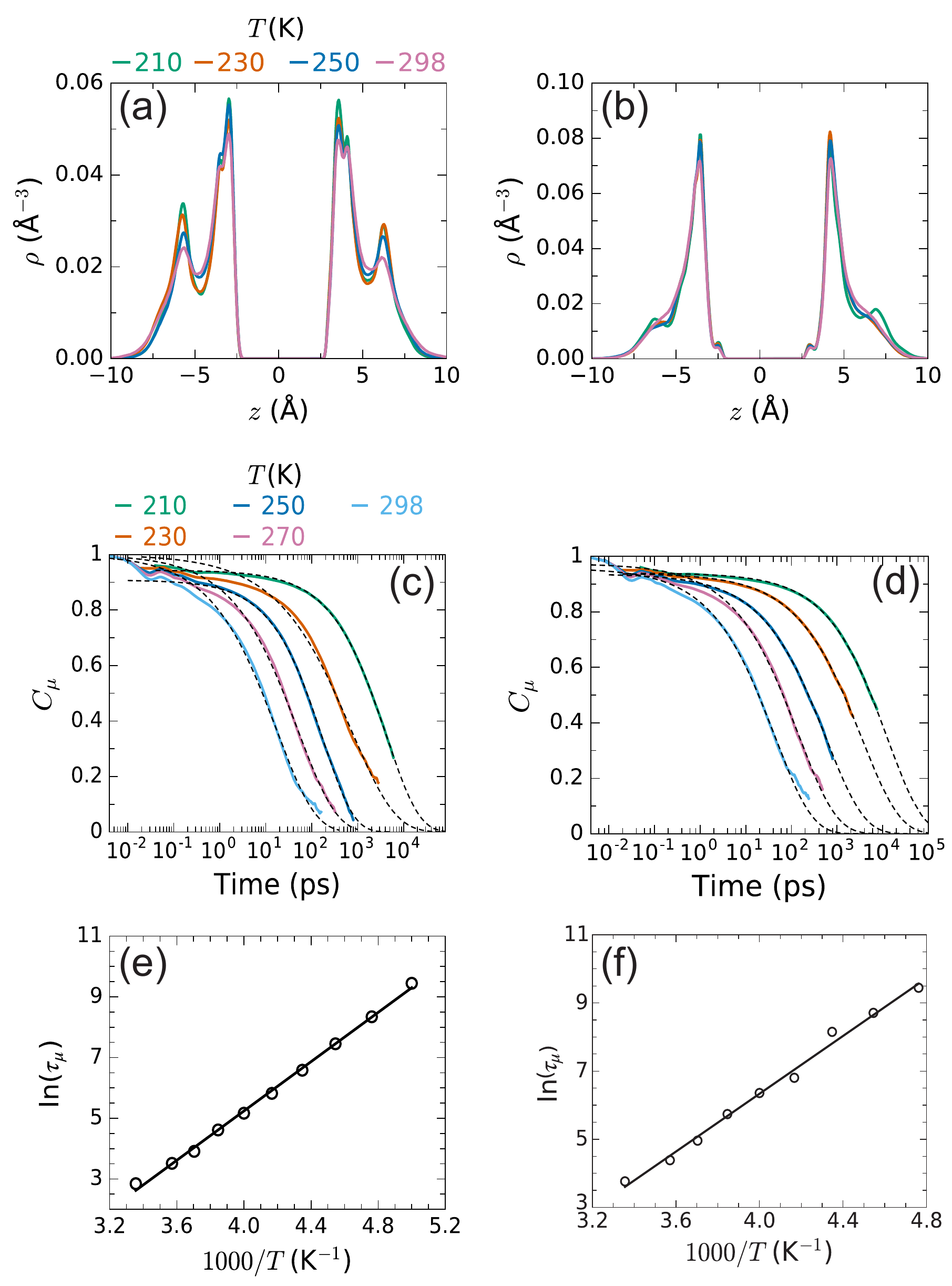}
	\caption[Dynamic crossover in dipole relaxation time of supercooled surface water]{Density distributions of surface water on (a) GO surface and (b) O surface, Dipole correlation function of surface water molecules on (c) GO surface and (d) O surface. The continuous lines are the MD simulation results, whereas dashed lines are fits obtained from Eq.~\ref{fit_ocf_IW}. The model parameter obtained from the fits of $C_{\mu}$ are given in Table~\ref{dipole_SE_SW}. Temperature dependence of the dipole relaxation time of surface water molecules on the (e) GO surface and (f) O surface. The open circles are simulation data and the solid lines are fits.}
	\label{DCF_SW}
\end{figure*}

In most molecular dynamics studies where the dynamics of supercooled water has been examined, a bulk water film co-exists with interfacial or bound water. In this scenario, the dynamics of interfacial water molecules is modulated by the presence of bulk water with which interfacial water molecules can form hydrogen bonds. To understand the extent of the influence of bulk water on the dynamic transitions of interfacial water, we have performed additional MD simulations at low hydration in the absence of  any bulk water, as illustrated in Fig.~\ref{snap_IW}b. As shown in Fig.~\ref{DCF_SW}a and b, the surface water molecules form layers (one or two) on the GO and O surfaces. We did not explore the G surface, which did not support a water layer due to the hydrophobic nature of the surface. Note that at low hydration, surface water is not in contact with bulk-like water. Since the density of water approaches zero after the first layer, we define the surface water region based on the local density, which is 30\% of the bulk density of water~\cite{monroe2018computational}. We compute the  dipole-dipole time correlation function (Fig.~\ref{DCF_SW}c-d) only for those surface water molecules that continuously reside on the surface. The dipole correlation functions are fitted to stretched exponential functions to extract dipole relaxation times, and model parameters are summarized in Table~\ref{dipole_SE_SW}. The temperature dependence of dipole relaxation time is illustrated in Fig.~\ref{DCF_SW}e and f for the surface water on GO and O surfaces. For surface water on the GO and O surfaces, the dipole relaxation time for the entire temperature range fits well with an Arrhenius equation with the activation energy of 33.75 and 35.26 kJ/mol for the GO and O surfaces, respectively. For the O surface, the activation energy for surface water is close to that of interfacial water. In the case of GO surface, however, the activation energy for surface water is less than that of interfacial water in the low-temperature region and higher than that of interfacial water in the high-temperature region. Interestingly,  surface water on the GO surface does not exhibit either an FTS or STS dynamic crossover in rotational motion (Fig.~\ref{DCF_SW}e and f), whereas the interfacial water showed a distinct STS transition. Surface water on the O surface did not depict a dynamical crossover either.

\begin{table}[!htpb]
	\caption[Model parameters for the dipole correlation function of supercooled surface water]{Model parameters obtained from the fitting of $C_{\mu}$ of surface water molecules on the GO and O surfaces at various temperatures. The $C_{\mu}$ is fitted to Eq.~\ref{fit_ocf_IW}.}
	\centering
	\begin{ruledtabular}
	\begin{tabular}{c c c c c c c}
		Surface & T(K) & t$_{w}$ (ps) & N$_{w}$ & a  & $\beta$ & $\uptau_{\mu}$ (ps)  \\	 
		\hline		
		GO & 200 & 9360 & 929 & 0.9258 & 0.7175 & 12670 \\ 
		& 210 & 8000 & 565 & 0.9437 & 0.6149 & 4202 \\
		& 220 &  7000 & 238 & 0.9837 & 0.5484 & 1726 \\
		& 230 &  3500 & 240 & 1.0 & 0.4553 & 772.6 \\
		& 240 &  1200 & 399 & 0.952  & 0.5597 & 337.2  \\
		& 250 &  1000 & 231 & 0.9081  & 0.6494 & 175.3 \\
		& 260 &  510 & 372 & 0.9182 & 0.6405 & 101  \\
		& 270 &  400 & 254 & 1.0 & 0.5182 & 49.91 \\
		& 280 &  300 & 315 & 0.9918 & 0.545 & 33.62  \\
		& 298 &  200 & 249 & 1.0  & 0.5165 & 17.21   \\		
		\\
		O  & 210 & 10000 & 597 & 0.9398  & 0.5602 & 12650 \\
		& 220 &  7000 & 363 & 0.9398 & 0.5212  & 6054  \\
		& 230 &  3150 & 414 & 0.9386 & 0.5188 & 3469 \\
		& 240 &  1500 & 525 & 0.9644  & 0.4823 & 900.4 \\
		& 250 &  1000 & 415 & 0.9367 & 0.5228 & 576.1 \\
		& 260 &  800 & 229 & 0.9636 & 0.5077 & 310.5  \\
		& 270 &  600 & 156 & 0.9736 & 0.5212 & 142.3 \\
		& 280 &  400 & 264 & 0.9579 & 0.5277 & 80.17  \\
		& 298 &  300 & 157 & 0.9577 & 0.5285 & 43.08   \\ 						
	\end{tabular}
	\end{ruledtabular}
	\label{dipole_SE_SW}
\end{table}



\begin{figure*}[!htpb]
	\centering
	\includegraphics[scale=0.65]{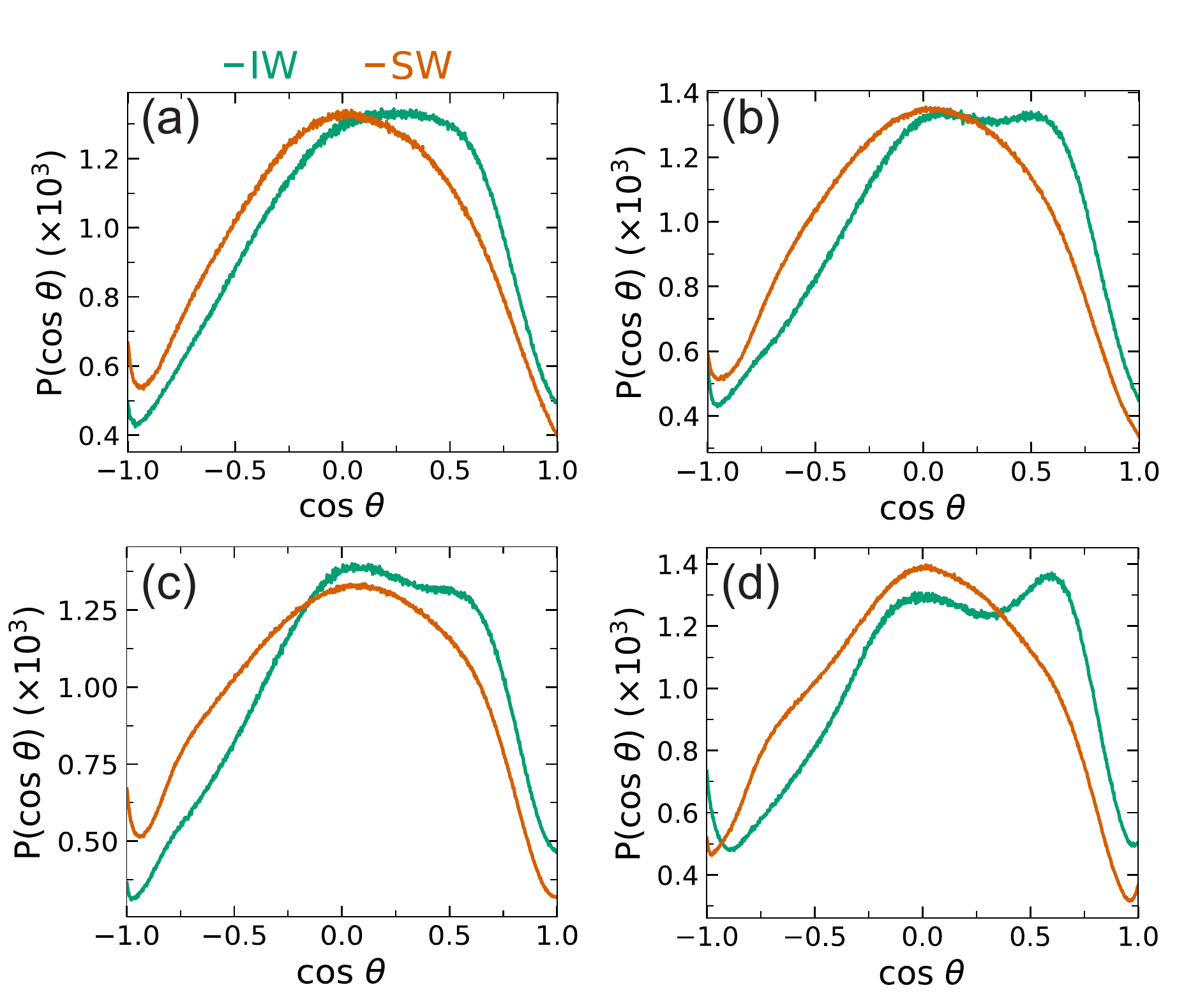}
	\caption[Comparison of dipole angle distribution of interfacial water and surface water on GO]{Comparison of dipole angle distribution of interfacial water (IW) and surface water (SW) on GO surface (a) $T$ = 298 K (b) $T$ = 260 K, (c) $T$ = 230 K, and (d) $T$ = 210 K}
	\label{angle_IW_SW_GO}
\end{figure*}

\begin{figure*} [!htpb] 
	\centering
	\includegraphics[scale=0.65]{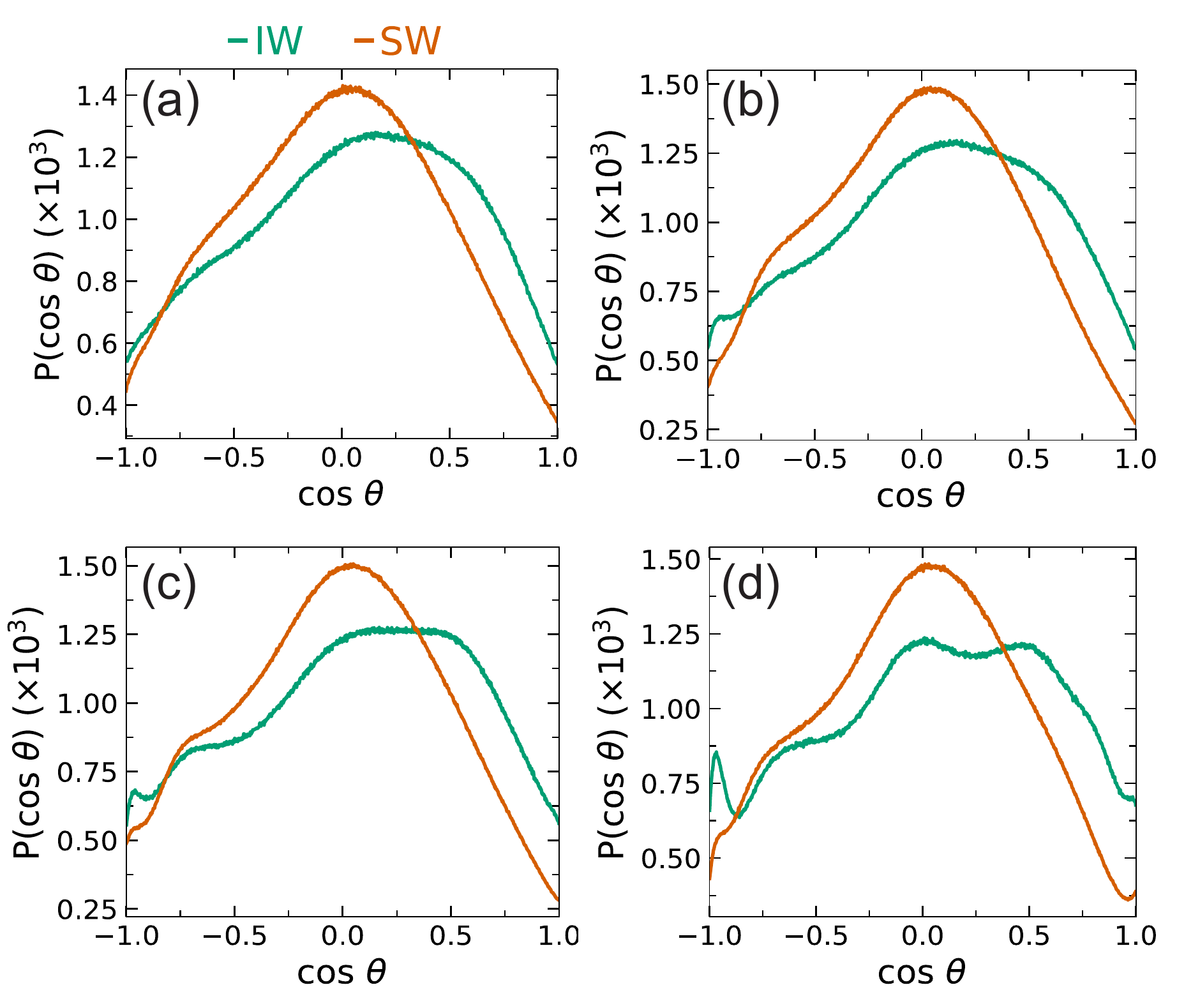}
	\caption[Comparison of dipole angle distribution of interfacial water and surface water on O]{Comparison of dipole angle distribution of interfacial water and surface water on O surfaces (a) $T$ = 298 K (b) $T$ = 260 K, (c) $T$ = 230 K, and (d) $T$ = 210 K. The dipole angle is angle formed by dipole moment of a water molecule with the surface normal.}
	\label{angle_IW_SW_O}
\end{figure*}

\section{Discussion and Conclusion}
We use molecular dynamics simulations to study the influence of surface chemistry and texture on the rotational relaxation dynamics of hydration water. In one case, a bulk water film is present adjacent to the surface, creating a situation where the hydration or bound water is always in contact with neighbouring bulk water molecules. In the second case, bulk water is absent, and only bound  water is present. We refer to these distinct water situations as interfacial and surface water, respectively. Table~\ref{data_summary} summarizes the dynamic crossover in the dipole relaxation time of interfacial and surface water on the GO, O, and G surfaces.

\begin{table}[!htpb]	
	\centering
	\caption{Summary of dynamic crossover in dipole relaxation of interfacial and surface water on GO, O and G surfaces}
	\begin{ruledtabular}
	\begin{tabular} {c c c c c c}		
System & Surface &  $T$ (K) & Fitting & $E_{\text{A}}$ (kJ/mol) & $T_{\text{C}}$ (K) \\ 
	\hline
		 & GO &  298 - 240 & Arrhenius & 28.28 &  \\
     & GO &  230 - 210 & Arrhenius & 47.51 & 230 \\
		\\
IW	 & O & 298 - 210 & Arrhenius & 35.39 &  \\
		\\
   & G & 298 - 240 & VFT &  &  \\
		 & G &  230 - 210 & Arrhenius & 21.17 & 232 \\    
\\
\\
		  & GO & 298 - 200 & Arrhenius & 33.75 &  \\
	SW	& O & 298 - 210 & Arrhenius & 35.26 &  \\			
	\end{tabular}
	\end{ruledtabular}
	\label{data_summary}
\end{table}

Our study reveals that the dynamic transition for the interfacial water is a strong function of the surface chemistry. For the GO surface, which is a combination of hydrophilic and hydrophobic regions, a STS transition is observed upon supercooling at a crossover temperature, $T_{\text{c}}$ of 230 K which lies 22 degrees below the melting point  (252 K) for the TIP4P/2005 water model used in this study. In the only other reported MD study of rotational relaxation~\cite{biswal2009dynamical}, an STS transition is observed for water in both the major and minor grooves of DNA where the crossover temperature was 19 degrees below the melting point for the TIP5P model used in their study. In the presence of a homogeneously hydrophilic surface created with a fully oxidized graphene, such as the  O surface used in this study, only  a single Arrhenius regime is observed over the temperature range of 298 - 210 K spanning temperatures well within the deeply supercooled regime. For the purely hydrophobic graphene surface, we observe a distinct FTS transition at a $T_{\text{c}}$ of 232 K, which is similar to the crossover temperature for the STS transition on the GO surface. 

With the exception of the work by Biswal et~al.~\cite{biswal2009dynamical} where the rotational relaxations have been investigated for hydration water, most MD studies of hydration water are related to the translational relaxation dynamics of hydration water, where comparisons are made with QENS results. The differences between the surfaces examined in this study offer insights into the rotational relaxation dynamics of hydration water. In general, bulk water has been observed to undergo an FTS transition, and this has been observed in MD simulations with a wide variety of water models~\cite{de2016mode}. In particular, the rotational relaxation of the TIP4P/2005 water model shows an FTS transition upon supercooling at a temperature of 214 K, revealing a rather strong dependence on density~\cite{de2016mode}. The fragility of water reveals the presence of more fluid-like states and the crossover temperature is the temperature that separates two dominant relaxation processes. At higher temperatures in the "fragile" regime, relaxation occurs via cage formation and breaking events, which allow molecules to translate and explore phase space. In the "strong" regime, relaxation is dominated by rarer hopping events and the fluid is considered to be in a more structurally arrested state. The presence of a continuous surface of hydrogen bonding sites on the O surface results in greater ordering and structural arrest of the interfacial water. This is the largest perturbation a water molecule experiences when compared with the environment in bulk water. As a consequence, interfacial water is seen to behave as a "strong" liquid well into the supercooled regime and no transition in rotational relaxation dynamics is observed up to 210 K. 

In the presence of the hydrophobic G surface, interfacial water has the least amount of hydrogen bonding and the rotational relaxation times are the lowest when compared with either the O or GO surface (see Table~\ref{dipole_SE_IW}). Additional signatures of greater mobility for the G surfaces are observed in the MSD data (Fig.~\ref{msd_IW}c). Note that this greater mobility and smaller relaxation times is despite the higher density of interfacial water for the G surface (Fig.~\ref{density_GO_LL}c) indicating that although water is able to form a high-density interfacial region dominated by van der Waals interaction with the surface, the lowered hydrogen bonding with the surface reduces the structural arrest when compared with either the O or GO surface. Thus interfacial water on the G surface has the closest resemblance to the dynamical transition in bulk water, showing a fragile regime at higher temperatures with a transition to a strong regime at lower temperatures. The dynamics of interfacial water in the GO surface lies between the two extremes observed in the O and G surfaces. The presence of hydrophilic regions covering 50\% of the surface in the GO surface is sufficient to impede the dynamics resulting in the observed STS transition. 

An important finding in our study is the manner in which bulk water modulates the relaxation dynamics. 
In the absence of bulk water, the relaxation dynamics of surface water on the  O surface shows a single Arrhenius behaviour similar to that observed with interfacial water. Similar activation energies for both situations suggest that surface functionalization dictates the rotational relaxation, with bulk water having a weaker influence on the rotational energy landscape. For the GO surface, however, the STS transition for interfacial water is reduced to a single Arrhenius behavior for surface water. This indicates that the presence of bulk water in the partially oxidized GO surface plays a stronger role in modulating the rotational energy landscape, suggesting that  water in the  hydrophobic graphene regions is influenced by the presence of bulk water to a greater extent when compared with water on a fully oxidized surface. 
In general, the dipole relaxation time of interfacial water on the  GO surface is larger than the surface water on the same surface. However, in the case of the O surface, the relaxation times are quite similar (Tables~\ref{dipole_SE_IW} and \ref{dipole_SE_SW}), indicating that bulk water can couple to a greater extent and alter the rotational relaxation of water on the  partly hydrophobic GO surface.  

Additional molecular insights can be obtained from the analysis of the dipole angle distributions for interfacial and surface water on different surfaces (Fig.~\ref{angle_IW_SW_GO}). For  surface water on the  GO surface, dipole angle distributions show more or less similar trends in the entire temperature range. However, the dipole orientation of  interfacial water on the GO surface changes to a weak but distinct bimodal distribution as the temperature is reduced with water dipoles sampling orientations where hydrogen atoms are directed away from the surface ($\cos \theta$ $ >$ 0). In contrast, the distribution for surface water is shifted toward orientations where hydrogen atoms are directed toward the surface ($\cos \theta$ $<$ 0) due to the absence of bulk water hydrogen bond acceptors. We tenuously link this distinct emergence of interfacial water bimodal orientational populations upon supercooling, to the STS transition observed for interfacial water in the GO system.  In contrast to the bimodal distribution observed for interfacial water on the GO surface, the dipole angle distribution for surface water, as well as for interfacial water, does not show a significant change in the entire temperature range for the O surface.  In the case of the O surfaces (Fig.~\ref{angle_IW_SW_O}), interfacial water has a broader distribution at lower temperatures, indicating that water dipoles sample a less restricted set of angles with respect to the surface resulting in a rotational energy landscape that is determined by a single Arrhenius behavior upon supercooling. A weak bimodal dipole distribution emerges for interfacial water on the O surface at the lowest temperature of $T$ = 210 K (Fig.~\ref{angle_IW_SW_O}d).

We finally comment on the water models and their influence on the results. While studying the effects of supercooling on the dynamic relaxation of water molecules, several water models are widely used in the literature. Although water models such as SPC/E and SPC accurately predict liquid-vapor coexistence and dynamic properties, the freezing points are significantly depressed; 215 K for the SPC/E model and 190 K for the SPC model. In studies of hydration water of lysozyme using the SPC/E model~\cite{camisasca2016two}, an FTS transition was observed in the translational relaxation time at $T_{\text{c}}$ = 215 $\pm$ 5 K which is at the freezing point for the water model, indicating the presence of a transition in the weakly supercooled regime. Studies on the translational relaxation dynamics of water in the major and minor grooves of DNA with the TIP5P model ($T_{m}$ = 274 K) illustrate the presence of an FTS transition at 255 K for water in the major grooves and an STS transition for water in the minor grooves.
In contrast, the $\uptau_{\mu}$ showed an STS transition for water in both the major and minor grooves of DNA at 255 K, which lies in the supercooled regime for this particular water model. The authors point out that rotational dynamics captures the local changes in the environment more faithfully when compared with the translational dynamics. Our observation of the STS transition for interfacial water in the GO surfaces and the "strong" nature of interfacial water on the O surface is consistent with this notion. Although every model has its advantages and disadvantages, our choice for the TIP4P/2005 water model is based on findings that this model provides an optimal balance between various predictive properties expected of rigid water models. Further, the temperature of maximum density for the TIP4P/2005 model has been found to agree remarkably well with experimental data~\cite{gallo2012mode}. 

In conclusion, our study reveals that the nature of the surface can significantly alter the relaxation dynamics of water when compared with the features of bulk water. Although it has been well known that surfaces play an important role in modulating structure and dynamics, our study shows for the first time the important role played by surface chemistry as well as the role of a co-existing bulk water film in modulating the relaxation dynamics of water in the supercooled regime.

\begin{acknowledgements}
We would like to thank the Supercomputer Education and Research Centre (SERC) and the Thematic Unit of Excellence (TUE) in the Solid State and Structural Chemistry Unit (SSCU) for computational resources and the Department of Science and Technology, India for funding.
\end{acknowledgements}

\bibliographystyle{apsrev4-2}
\bibliography{Reference_IW} 

\end{document}